\documentclass[twocolappendix]{emulateapj}

\usepackage[usenames,dvipsnames]{color}
\newcommand{\hilight}[1]{{#1}}
\usepackage{comment}

\shorttitle{Top-Heavy GRB Jets} \shortauthors{Duffell \& MacFadyen}

\begin{document}

\title{From Engine to Afterglow: Collapsars Naturally Produce Top-Heavy
Jets and Early-Time Plateaus in Gamma Ray Burst Afterglows}

\author{Paul C. Duffell and Andrew I. MacFadyen} \affil{Center for
Cosmology and Particle Physics, New York University}
\email{pcd233@nyu.edu, macfadyen@nyu.edu}

\begin{abstract}

We demonstrate that the steep decay and long plateau in the early phases
of gamma ray burst (GRB) X-ray afterglows are naturally produced in the
collapsar model, by a means ultimately related to the dynamics of relativistic jet propagation through a massive star. We present two-dimensional axisymmetric hydrodynamical simulations which start from a collapsar engine and evolve all the way through the late afterglow phase.  The resultant outflow includes a jet core which is
highly relativistic after breaking out of the star, but becomes
baryon-loaded after colliding with a massive outer
shell, corresponding to mass from the stellar atmosphere of the
progenitor star which became trapped in front of the jet core at
breakout.  The prompt emission produced before or during this collision
would then have the signature of a high Lorentz factor jet, but the
afterglow is produced by the amalgamated post-collision ejecta which has
more inertia than the original highly relativistic jet core and thus has
a delayed deceleration.  This naturally explains the early light curve
behavior discovered by Swift, including a steep decay and a long
plateau, without invoking late-time energy injection from the central
engine.  The numerical simulation is performed continuously from engine
to afterglow, covering a dynamic range of over ten orders of magnitude
in radius. Light curves calculated from the numerical output
demonstrate that this mechanism reproduces basic features seen in early
afterglow data. Initial steep decays are produced by internal shocks,
and the plateau corresponds to the coasting phase of the outflow.

\end{abstract}


\section{Introduction} \label{sec:intro}
%
%1992ApJ...392L...9D,
%1992Natur.357..472U, 1998AnA...333L..87D, 2001ApJ...552L..35Z,
%2014arXiv1404.0283V

Possibly the most important result of \textit{Swift} observations is the
discovery of early-time plateaus in GRB afterglows
\citep{2004ApJ...611.1005G, 2007RMxAC..27..140G}.  There have been many scenarios invoked to explain this plateau.  One explanation requires late-time energy injection from the GRB central engine \citep{2006ApJ...642..389N, 2006ApJ...642..354Z, 2014arXiv1402.5162V}. Another proposition is that the radiation is produced in a long-lived reverse
shock \citep{2007ApJ...665L..93U, 2007MNRAS.381..732G}.  Other ideas
include evolving microphysics \citep{2006MNRAS.369.2059P, 2006MNRAS.370.1946G} or a slow
energy transfer from ejecta to the circumburst medium \citep{2006ApJ...642..389N, 2006MNRAS.366L..13G, 2007ApJ...655..973K}.  It has also been suggested that viewing-angle effects may be responsible \citep{2006ApJ...641L...5E, 2007RMxAC..27..140G}.  Another possibility is that the plateau
corresponds to the time before the deceleration phase of the afterglow
\citep{2011ChPhL..28l9801L, 2012ApJ...744...36S}, however this requires
very modest Lorentz factors $\gamma \sim 30$.  Other popular jet models
which attempt to explain afterglow plateaus are two-component jet models
\citep{2002MNRAS.337.1349R, 2003Natur.426..154B, 2005ApJ...626..966P,
2006MNRAS.370.1946G, 2011AnA...526A.113F}.  The usual idea of two-component flows assumes
that the afterglow is a combination of emission from a narrow highly
relativistic jet core and a wide, less relativistic envelope.

\begin{figure*} 
\epsscale{1.0} 
\plotone{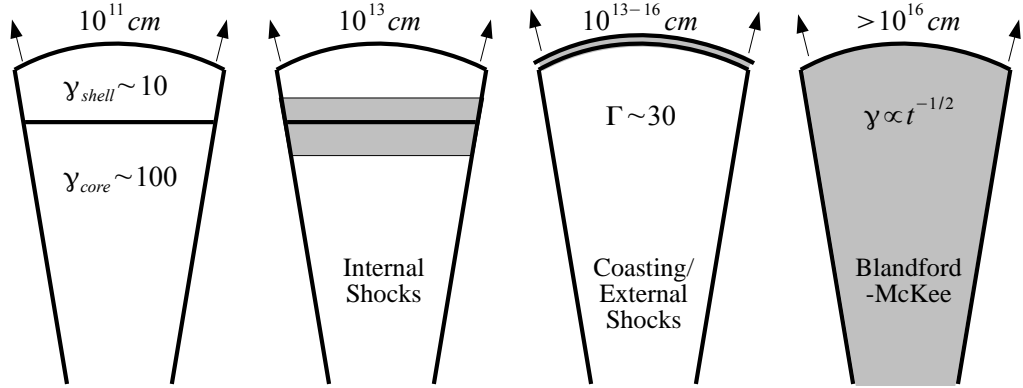} 
\caption{ 
Various stages of the top-heavy jet are depicted in a cartoon illustration.  The
jet core is clean but the outer shell is loaded with baryons, resulting
in a much lower Lorentz factor than the jet core.  These two components
eventually collide, producing internal shocks and amalgamating into a
single jet with reduced Lorentz factor $\Gamma \sim 30$.  The grey
regions in the figure indicate shocked gas. 
\label{fig:cartoon} }
\end{figure*}

In this work, the goal is not to propose a given jet model and calculate
the resultant light curve.  Rather, a numerical calculation is performed
starting from a collapsar engine \citep{1993ApJ...405..273W, 1999ApJ...524..262M, 2001ApJ...550..410M} and evolving through the late afterglow
phases, to determine what outflows naturally look like, and what sort of
light curves are generically produced.  The major discovery of this work
is that light curves broadly similar to those discovered by \textit{Swift} are
naturally produced without any special ingredients, other than a
collapsar which initially collimates the flow.

The outflow produced by the collapsar is not a standard two-component
jet, but a ``top-heavy'' jet, which has radial as well as angular
variation.  In this case, a highly relativistic jet core is preceded by
a heavy, baryon-loaded outer shell to the jet.  The shell is massive
enough to decelerate the entire jet core significantly upon collision. 
This collision produces internal shocks which will produce synchrotron emission, and the final result of the collision is a less
relativistic amalgamated jet, which is responsible for the afterglow
emission (see Figure \ref{fig:cartoon}).

The main goals of this study are to demonstrate that such a top-heavy
jet is a natural outcome of a collapsar GRB engine, and that this
top-heavy flow naturally produces the basic features of the early
afterglow light curves, including long plateaus.  The jet core consists
of the relativistic material ejected from the engine after the jet has
tunneled through the stellar interior.  The heavy outer shell of the jet
consists of material from the progenitor atmosphere which was originally
in front of the jet head at breakout.

All of this is demonstrated numerically, in the first multidimensional
numerical GRB study which evolves the entire burst, from engine to
afterglow.  This covers a dynamic range of over ten orders of magnitude
in length scale.  To date, such a range of scales has only been
numerically resolved in one-dimensional studies, e.g.
\cite{1999ApJ...513..669K}.

\section{Numerical Set-Up} \label{sec:numerics}

The system is assumed to be governed by the equations of relativistic hydrodynamics,

\begin{equation} \partial_{\mu} ( \rho u^{\mu} ) = S_D \label{eqn:claw1}
\end{equation} \begin{equation} \partial_{\mu} ( ( \rho + \epsilon + P )
u^{\mu} u^{\nu} + P g^{\mu \nu} ) = S^{\nu} \label{eqn:claw2}
\end{equation} 

where $\rho$ is proper density, $P$ is pressure,
$\epsilon$ is the internal energy density, and $u^{\mu}$ is the
four-velocity.  The equations are solved in two dimensions assuming axisymmetry.  Magnetic fields have also been ignored in this calculation, so that this is a purely hydrodynamical jet model.  The source terms $S_D$ and $S^{\nu}$ model the engine,
which injects mass, energy and momentum into the system at small scales.
 An adiabatic equation of state is employed,
 
\begin{equation}
P = (\hat \gamma - 1) \epsilon
\end{equation}

with adiabatic index $\hat \gamma = 4/3$, and relativistic units are
chosen such that $c = 1$.

The hydrodynamical evolution is carried out using the JET code
\citep{2011ApJS..197...15D, 2013ApJ...775...87D, 2014arXiv1403.6895D}, a
moving-mesh technique tailored to radial outflows.  This numerical
method is particularly important for this problem, both because of the
high Lorentz factors $\gamma \sim 100$ which must be fully resolved, and
the tremendous range of length scales covered.  In particular, the inner
and outer boundaries are moved during the calculation, so that it is
possible to evolve the jet over many orders of magnitude in radius,
while only resolving a single order of magnitude at any one time (see Figure \ref{fig:radius}).

\subsection{Initial Conditions}

\begin{figure} 
\epsscale{1.0} 
\plotone{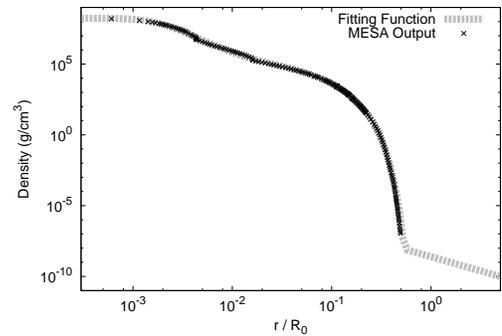} 
\caption{ 
Initial conditions for this calculation are determined from the output of the stellar evolution code MESA.  A fitting function is used (Equation \ref{eqn:stmodel}) in order to make the results as reproducible as possible.  The fitting function is plotted here alongside the MESA output, demonstrating a reasonable fit to the stellar model.
\label{fig:mesa} }
\end{figure}

The initial set-up is designed to model a jet breaking out of a
Wolf-Rayet star.  The initial data for this star was acquired using the
MESA code \citep{2011ApJS..192....3P, 2013ApJS..208....4P} to evolve a
stellar model until just before core collapse.  This provides density as
a function of radius which will affect the dynamics of the jet and
cocoon as the jet tunnels its way through the star.

The initial stellar model assumes a low-metallicity rapidly-rotating
star, with an initial mass of $30 M_\sun$ and rotating at $99$\% of
breakup velocity.  The star exhibits significant mass loss during its
evolution, and the final state of the star before collapse is a
Wolf-Rayet star with a mass of $18 M_\sun$ and about half the radius of
the sun.  The choice of rotation velocity affects the initial mass and radius of the progenitor, but it was chosen assuming that a rapidly rotating progenitor may be necessary for jet production.  In order to provide a standardized stellar model, a fitting
function is used to approximate the result of the MESA output.  The
density as a function of radius is well-fit by the following:

\begin{equation} \begin{array}{rcl} \rho( r , 0 ) & = & {\rho_c (
\text{max}(1 - r/R_3,0) )^n \over 1 + (r/R_1)^{k_1}/( 1 + (r/R_2)^{k_2}
) } \\ \\ & & + \rho_{\rm wind} (r/R_3)^{-2} \end{array} \label{eqn:stmodel}
\end{equation}

All of these parameters are listed in Table \ref{tab:engine}.  The fitting function and MESA output are shown in Figure \ref{fig:mesa}.  Note that outside the stellar radius, the density has a wind profile $\rho(r) = A/r^2$, where $A = \rho_{\rm wind} R_3^2 = 1.2 \times 10^{13}$ g/cm $\approx 24$ times the typically assumed value of $5 \times 10^{11} g/cm$, which assumes a wind velocity of $1000$ km/s and a mass loss rate of $10^{-5} M_{\sun}$ per year.  The density surrounding the progenitor at death is not well known, and a factor of $24$ above this fiducial value is not outside of the range of reasonable values for this constant.  Velocity and pressure are initially set to negligible values:

\begin{equation} \vec v(r,0) = 0, \end{equation}

\begin{equation} P(r,0) = 10^{-6} \rho(r,0). \end{equation}

Of course, the injected jet will quickly raise these to non-negligible values in its vicinity.  Gravity is not included in this calculation due to the short engine
timescale considered; all of the dynamics are caused by the injection of
energy and momentum at small radii.

\subsection{Engine Model}

The jet is injected deep within the stellar interior.  The true engine operates on unresolved scales and consists of poorly understood physics.  In the current study, the engine will be injected on resolved scales using a parameterized model.  This is accomplished by the use of source terms in the hydrodynamical
equations (\ref{eqn:claw1}, \ref{eqn:claw2}).  The engine is
parameterized by a power, Lorentz factor, baryon loading, injection
angle, injection radius, and engine duration (see Table
\ref{tab:engine}).

\begin{table} \caption{Stellar and Engine Parameters} \label{tab:engine}
\begin{center} \leavevmode \begin{tabular}{lll} \hline \hline Variable  
& Definition              & Value          \\ \hline 
$M_0$         & Characteristic Mass Scale   & $2 \times 10^{33}$ g  \\ 
$R_0$         & Characteristic Length Scale & $7 \times 10^{10}$ cm  \\ 
$\rho_c$      & Central Density             & $3 \times 10^{7} M_{0}/R_0^3$  \\ 
$R_1$         & First Break Radius          & $0.0017 R_0$  \\ 
$R_2$         & Second Break Radius         & $0.0125 R_0$  \\ 
$R_3$         & Outer Radius                & $0.65 R_0$ \\ 
$k_1$          & First Break Slope          & 3.24  \\ 
$k_2$          & Second Break Slope         & 2.57 \\ 
$n$            & Atmosphere Cutoff Slope    & 16.7  \\
$\rho_{\rm wind}$  & Wind Density               & $10^{-9} M_{0}/R_0^3$  \\
\hline 
$\theta_{0}$   & Injection Angle            & 0.1  \\
$\gamma_{0}$   & Injected Lorentz Factor    & 50  \\ 
$\eta_{0}$     & Energy-to-Mass Ratio       & 100 \\ 
$r_{0}$        & Nozzle Size                & 0.01 $R_0$  \\ 
$L_{0}$        & Engine Power (One-Sided)   & $2 \times 10^{-3} M_{0} c^3/R_0$  \\ 
               &                            & $=1.5 \times 10^{51}$ erg/s  \\ 
$\tau_{0}$     & Engine Duration            & 4.3 $R_0/c$  \\ 
               &                            & $= 10$ seconds  \\
\hline \multicolumn{3}{l}{}                                            
\\ \end{tabular} \end{center} \end{table}

\begin{figure*} \epsscale{1.0} 
\plotone{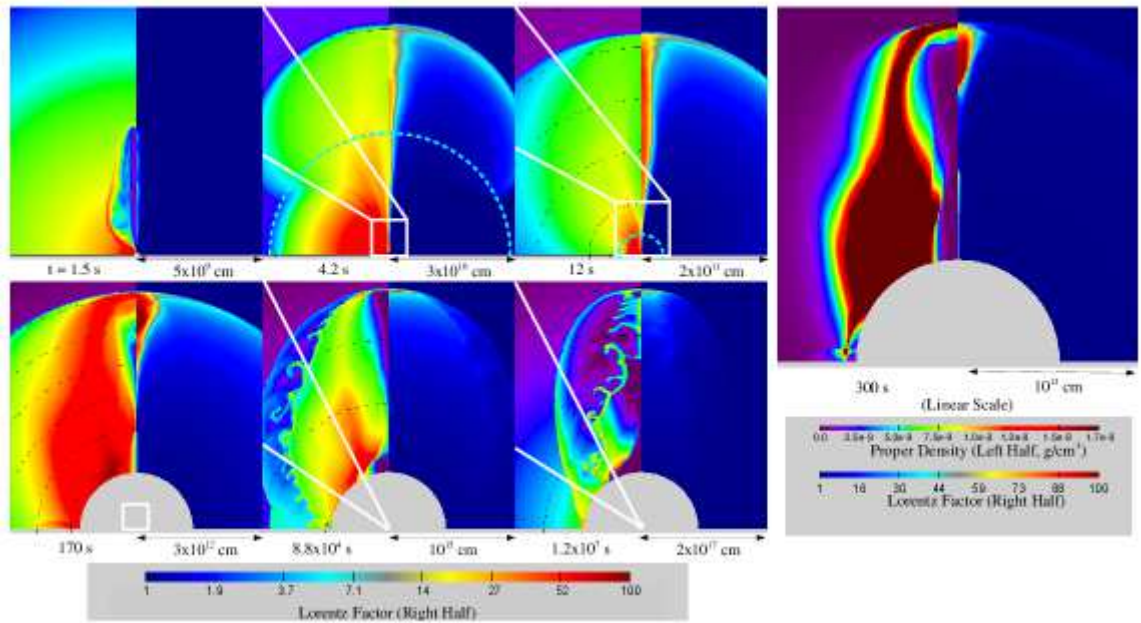} \caption{ Snapshots
showing the evolution of the jet from the engine deep in the interior of
the star to breakout from the star to the collimated outflow which
produces afterglow emission.  Overall, the calculation spans over ten
orders of magnitude in radius.  The left half of each panel displays
logarithm of density, and the right half is logarithm of the Lorentz
factor.  The large panel on the right is a snapshot at $t = 300$ seconds, shown in linear scale so that the massive shell is much more prominently displayed.  In the top-center and top-right small panels ($t=4.2$ s and $t=12$ s), the thick dashed cyan curves indicate the surface of the progenitor, showing clearly that the jet breaks out before the engine turns off at $t=10$s.\label{fig:pretty} } 
\end{figure*}

The source terms are expressed in terms of the nozzle function,
$g(r,\theta)$:

\begin{equation} g(r,\theta) \equiv (r/r_0) e^{-(r/r_0)^2 /2}e^{ (cos
\theta - 1) /\theta_0^2 }/N_0 \end{equation}

where $N_0$ is the normalization of $g$:

\begin{equation} N_0 \equiv 4 \pi r_0^3 ( 1. - e^{-2/\theta_0^2} )
\theta_0^2 \end{equation}

The source terms in Equations (\ref{eqn:claw1}) and (\ref{eqn:claw2})
are given by the following:

%\begin{equation} %S^0 = L_0 e^{-t/\tau_0} g(r,\theta), %\end{equation}
%
%\begin{equation} %S^r = S^0 \sqrt{ 1 - 1/\gamma_0^2 }, %\end{equation}
%
%\begin{equation} %S_D = S^0/\eta_0. %\end{equation}
%
\begin{eqnarray} 
S^0 & = & L_0 e^{-t/\tau_0} g(r,\theta), \\
S^r & = & S^0 \sqrt{ 1 - 1/\gamma_0^2 }, \\ 
S_D & = & S^0/\eta_0. 
\end{eqnarray}

The parameters in the above equations for the source terms are all
listed in Table \ref{tab:engine}.  Engine power and duration are taken as typical values inferred from observations.  The rest of the engine parameters are somewhat unconstrained, so the parameters are chosen so as to produce a reasonably collimated relativistic outflow.  This model has many free parameters which could be varied to produce different jet properties.  This has been done, for example, by \cite{2001ApJ...550..410M, 2003ApJ...586..356Z, 2004ApJ...608..365Z, 2007ApJ...665..569M}.  It should also be noted that most jet
breakout calculations use a ``nozzle" boundary condition instead of a
source term to model the engine.  Either way this models unresolved
physics, but the source-term method appears to produce more numerically stable outflows.  As mentioned above, the injection radius $r_0$ is much larger than the true engine.  This was done so as to ensure good resolution of the engine, which is covered by hundreds of computational zones, $r_0 / \Delta r_{\rm min} \sim 10^3$.  The size of the engine may impact the flow ejected from the progenitor.

\subsection{Afterglow Light Curves}

A synchrotron emission model is used to construct an afterglow light
curve from the hydrodynamical output.  The model is based on the one
employed in \cite{2010ApJ...722..235V}, though it is simplified in that
it assumes the flow is optically thin.  Electron cooling is accounted
for using a global cooling timescale.  The jet is also assumed to point
directly at the observer.  The synchrotron model parameters are
summarized in Table \ref{tab:synch}.  Note that this synchrotron model is designed for X-ray afterglows, and is less reliable in lower-frequency bands due to the global cooling timescale.  All current results are presented in the X-ray, and multi-band studies are left for a future investigation which will employ a more complete synchrotron model.

\section{Results} \label{sec:results}

Figure \ref{fig:pretty} shows several stages in the evolution of the
jet.  First the engine drills out a tunnel in the stellar interior,
supported by the hot cocoon surrounding it.  The jet eventually breaks
out of the star, producing a top-heavy outflow.  The outer shell breaks
out with a modest Lorentz factor ($\gamma_{\rm shell} \sim 10$), in front of
a highly relativistic ($\gamma_{\rm core} \sim 100$) jet core.  \hilight{The thickness of this shell can be measured using data from the upper center panel of Figure 3.  The thickness is calculated to be $\sim 10^{10}$ cm; analytic calculations of \cite{2003ApJ...584..390W} would predict a cork thickness of order $\theta R_3 \sim 5 \times 10^9$ cm, in rough agreement with our results.}  Because the material is all moving at nearly the same speed ($c$), the jet core does not collide with the shell until it has expanded by several orders of magnitude ($t \sim \Delta r_{\rm shell} ~ \gamma_{\rm shell}^2$).  When the
core and shell collide, internal shocks are produced, and the two
components of the jet merge to a single jet with modest Lorentz factor
($\Gamma \sim 30$).  This process ends when the entirety of the jet core
is absorbed, at $t \sim \tau_0 \Gamma^2$.  Internal shocks are therefore produced until the outflow has expanded to a radius of roughly

\begin{equation}
R_{\rm internal ~shocks} \sim 10^{15} \text{~cm} \left( {\tau_0 \over 30 \text{~s} }\right) \left( {\Gamma \over 30} \right)^2
\end{equation}

\hilight{This radius of $10^{15}$ cm can be seen directly in Figure 5, as the transition from colliding to coasting occurs at $t \sim 3 \times 10^4$ seconds, when the flow has expanded to a radius of $ct \sim 10^{15}$ cm.}  This merged jet has more
inertia than the original jet core, and so it decelerates at a later
time than it would at the larger Lorentz factor $\gamma \sim 100$. 
Before this late deceleration time, there is a long coasting phase while
the jet sweeps up a negligible amount of mass from the external medium. 
This translates into a long plateau in the afterglow light curve.

\hilight{Figure 5 shows Lorentz factor as a function of time, measuring Lorentz factor in two different ways.  ``Maximum Lorentz factor" $\gamma_{\rm max}$ probes the fastest-moving material.  Before the collision, this probes the jet core, whereas post-collision $\gamma_{\rm max}$ probes the amalgamated jet.  ``Average Lorentz Factor" is weighted by energy, and essentially probes the average of the two components combined together.  Thus, there is only a modest change in $\gamma_{\rm avg}$ during the internal collisions, as the two flows amalgamate.}

\begin{table} \caption{Synchrotron Model Parameters} \label{tab:synch}
\begin{center} \leavevmode \begin{tabular}{lll} \hline \hline 
Variable      & Definition               & Value          \\ \hline 
$\nu_{\rm obs}$   & Observed Frequency       & $2.4 \times 10^{18}$ Hz  \\ 
$\epsilon_B$  & Magnetic Energy Fraction & $0.01$  \\ 
$\epsilon_e$   & Electron Energy Fraction & $0.1$  \\ 
$p$            & Synchrotron Power Spectrum Slope & $2.5$  \\
$d_L$          & Luminosity Distance      & $5.5 \times 10^{28}$ cm \\ 
$z$            & Redshift                & $2.2$ \\               
\end{tabular} \end{center} \end{table}

\begin{figure} 
\epsscale{1.0} 
\plotone{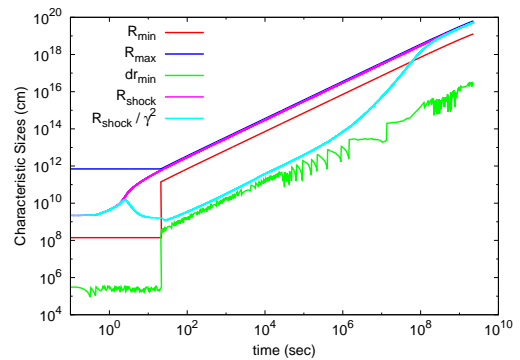} 
\caption{ Characteristic radii as a function of time, showing the motion of the inner and outer boundaries, $R_{\rm min}$ and $R_{\rm max}$, which are moved to follow the flow.  Also shown is the position of the blastwave, $R_{\rm shock}$, the characteristic width $R_{\rm shock}/\gamma^2$, and the smallest resolved scale $\Delta r_{\rm min}$.
\label{fig:radius} } 
\end{figure}

\begin{figure} 
\epsscale{1.0} 
\plotone{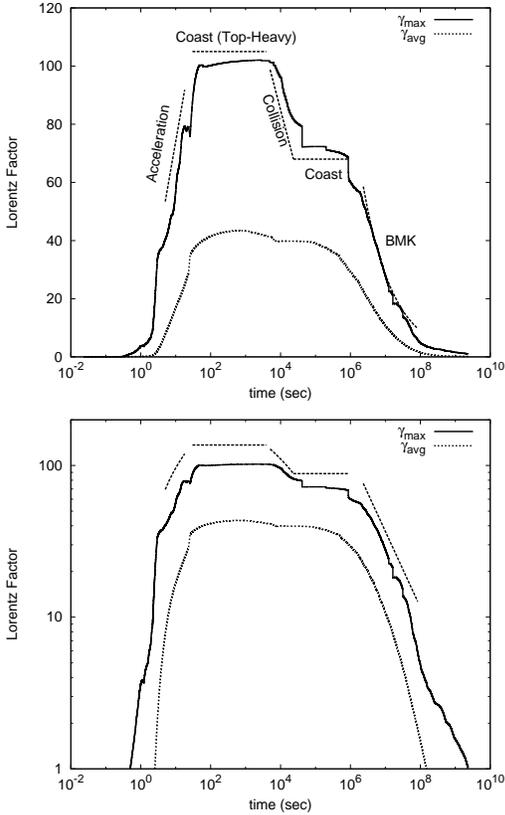} 
\caption{ Plotting the maximum Lorentz factor on the grid shows the sharp transition from the early coasting phase at $\gamma \sim 100$ to the amalgamated coasting at lower Lorentz factor.  Average Lorentz factor (weighted by energy) shows that this flow consistently has $\gamma \sim 40$ throughout this collision.
\label{fig:gamma} } 
\end{figure}

\begin{figure} 
\epsscale{1.0} 
\plotone{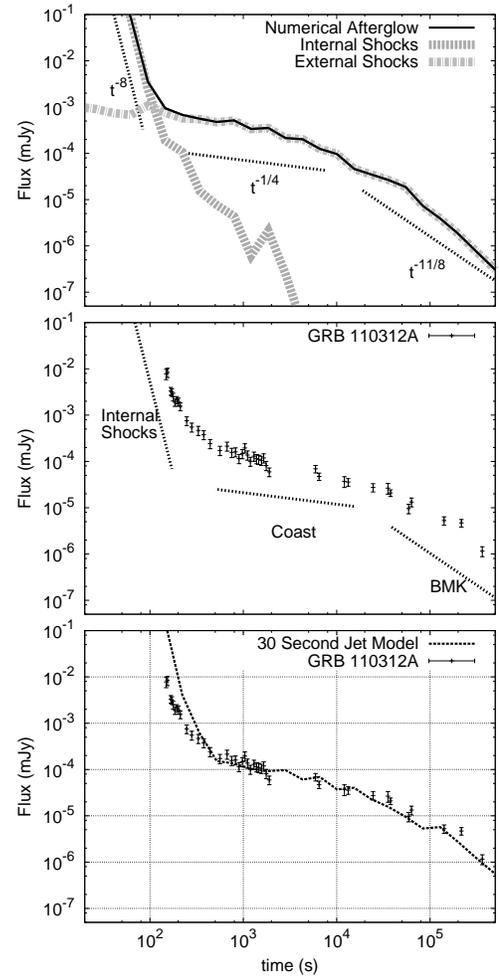} 
\caption{ X-Ray light
curves are produced from the numerical solution, assuming a synchrotron
radiation model.  A bright flare with steep decay until $t_{\rm obs} \sim
100$ seconds is produced by internal shocks, from the collision of the
jet core with the baryon-loaded shell ahead of it.  The deceleration
phase occurs after $t_{\rm obs} \sim 10^{4-5}$ seconds have passed, and
between the flare and deceleration there is a long plateau corresponding
to the coasting phase of the jet.  The afterglow model also uses a
passive scalar quantity to distinguish between emission from the ejecta
and emission from the circumburst medium.  This makes it possible to
tell the difference between emission from external and internal shocks,
shown in the figure.  Included in the second panel for comparison is
Swift X-Ray data from GRB 110312A.  This burst shows the same basic
shape and features as the afterglow model with the chosen parameters. 
In the third panel, it is demonstrated how one might fit the model
parameters to data, exploiting the scale invariance of the underlying
fluid equations.  Varying the characteristic scale $R_0$ provides a
better fit to data, which implies a larger progenitor $R_3 \approx 2
R_{\sun}$ and a longer-duration engine $\tau_0 \approx 33$ s.
\label{fig:afterglow} } 
\end{figure}

The light curve is shown in Figure \ref{fig:afterglow}.  Indicated in
this figure are the various stages in the evolution of the jet.  The
collision of the core with the outer shell produces internal shocks
which give rise to a flare which can last until observer time $t_{\rm obs} \sim 100$ seconds.  After this time there is a very steep decay,
followed by a long plateau over several orders of magnitude.  During
this time the jet is massive enough that it is not forced to decelerate
by the surrounding medium; this is the ``coasting phase" of the jet.  A
forward shock is present ahead of the flow at this time, but its
presence is not significant enough to affect the evolution of the jet
until enough mass has been swept up, at the late deceleration time
$t_{\rm obs} \sim 10^{4-5}$ seconds.  After this time, the jet begins to
decelerate and the light curve begins to exhibit a power-law dependence
consistent with the Blandford-McKee solution
\citep{1976PhFl...19.1130B, 1998ApJ...497L..17S}.

Also included in Figure \ref{fig:afterglow} is Swift X-Ray data from GRB
110312A \citep{2011GCNR..327....1O, 2011GCN..11786...1E}.  \hilight{The redshift of this burst is not known and is therefore treated as a free parameter.}  The plateau
phase is reasonably modeled by the coasting, $\Gamma \sim 30$ jet, as is
the late deceleration phase (consistent with the Blandford-McKee
solution in a wind circumburst medium).  The break times in this burst
also coincide reasonably well with the numerical model in this work. 
However, in this instance they are are about a factor of two later than
those in the considered progenitor model.  This suggests that this burst
would be better-fit by larger progenitors and longer-duration jets.

\subsection{Improving the Fit to Data}

It has clearly been demonstrated so far that the general shape of the
light curve agrees with an example afterglow dataset (Top two panels of
Figure \ref{fig:afterglow}).  This example was chosen partially because it is well-sampled, and partially because it fit well with the acquired light curve for these parameters, but it should be emphasized that most early afterglows exhibit this same characteristic structure.  It should also be emphasized that this required
no special ingredients; the set of parameters of the progenitor star and
engine was chosen at the beginning based on reasonable estimates, and
therefore this result might be generic, as it did not require detailed
tuning to achieve.

On the other hand, it is straightforward to use the inherent scale
invariance of the fluid equations to find a better fit to the data \citep{2012ApJ...747L..30V} (see also \cite{2012MNRAS.421.2610G}).  The
bottom panel in Figure \ref{fig:afterglow} shows an afterglow
constructed from the same numerical solution, but now assuming $R_0 =
2.3 \times 10^{11}$ cm.  For example, the progenitor radius is $R_3
\approx 2 R_{\sun}$ and the engine duration is $\tau_0 = 33$ seconds
(The energy of the burst and mass of the progenitor are still the same,
as $M_0$ has not been rescaled).  Not only does this provide a better
fit to the data, but encouragingly this is a first step toward learning
about the nature of the progenitor from the early stages of the afterglow light curve, by adjusting model parameters to fit the data.  A more comprehensive study would vary all model parameters to find a fit; for now we vary only the parameter $R_0$ as this is easy to do (without performing additional numerical runs) using the scale invariance of the fluid equations.  Future studies may be able to match afterglow data to model parameters by varying all of these parameters explicitly.

\begin{figure} 
\epsscale{1.0} 
\plotone{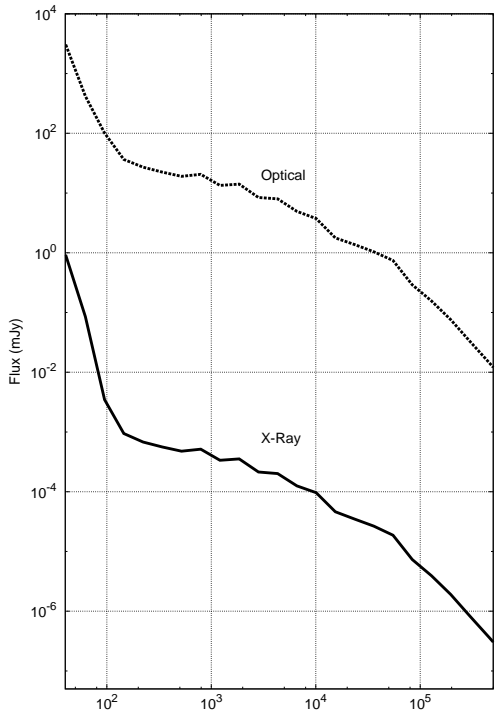} 
\caption{ An optical light curve is also calculated ($\nu = 5 \times 10^{14}$ Hz), and compared with X-Rays.  The plateau and deceleration phases are essentially identical as these frequencies are both above the cooling break during this time (optical is just a re-scaled version of X-rays with $F_{Op} = F_X (\nu_{Op}/\nu_X)^{-p/2}$).  However, at early times, the steep decay is less steep in the optical band.
\label{fig:xrayopt} } 
\end{figure}

\section{Discussion} \label{sec:disc}

This work has demonstrated the following: first, that GRB afterglow
plateaus are consistent with the coasting phase of a low Lorentz factor
jet, and secondly that such a jet naturally arises in a collapsar
scenario due to the amalgamation of the highly relativistic jet core
with a baryon-loaded outer shell.  Additionally, this work demonstrates
that internal shocks between the core and outer shell naturally produce
the steep decay found at early times. 

The details of this result are strongly dependent on the massive shell in front of the jet.  \hilight{During the time the jet is propagating through the star, there is a ``jet head" which is not trapped in front of the engine, but can be pushed aside.  However, as the jet accelerates and breaks out, there is a certain amount of material which becomes trapped in front of the jet \citep{2003ApJ...584..390W}.  The amount of material trapped in the shell is difficult to precisely calculate even numerically; in fact it is found to be less pronounced in 3D studies than in 2D \citep{2004ApJ...608..365Z}.}  This shell's properties will depend significantly on engine and progenitor properties, and on 3D and MHD effects.  This dependence has been explored numerically in  simulations of jet propagation through a stellar envelope
\citep{2001ApJ...550..410M, 2003ApJ...586..356Z, 2004ApJ...608..365Z, 2007ApJ...665..569M}.
Such an outer shell has also been referred to a ``cork'' and suggested
as source of precursor emission \citep{2003ApJ...584..390W} and as a
``breakout shell''and suggested as a facilitator of prompt emission
\citep{2006ApJ...651..333T}.  The idea that this shell might
decelerate the jet was also suggested by \cite{2012ApJ...744...36S}.  A similar idea was proposed to explain the radio afterglow of the giant flare from SGR1806-20 \citep{2006ApJ...638..391G}.  This explanation for afterglow plateaus has a large number of important consequences.

\subsection{Requirement of Wind near Progenitors}

If the plateau phase is described by a small negative slope, this
requires that the external density profile during this phase be
consistent with a stellar wind, $\rho(r) \propto r^{-2}$.  The slope of
the afterglow during this phase can be calculated straightforwardly
\citep[e.g.][]{2012ApJ...744...36S, 2005ApJ...631.1022G}.  During this time the ejecta pushes a forward shock ahead of it, one which is too weak to decelerate the
ejecta, but still powerful enough to produce significant synchrotron
radiation.  The scaling of the light curve can be calculated by modeling
the system as a piston with constant Lorentz factor $\Gamma$ pushing a
shock ahead of it at this same Lorentz factor.  Since $\Gamma$ is fixed,
it can be ignored in the scaling arguments.  The shock jump conditions
give

\begin{equation} P_{\rm shock} \propto \rho_{\rm ext}(r),~~ \rho_{\rm shock} \propto
\rho_{\rm ext}(r), \end{equation}

where $\rho_{\rm ext}(r) \propto r^{-k}$, with $k = 2$ for a wind.  Assuming
high frequencies in the slow cooling regime $\nu_m < \nu_c < \nu$ (which
is true in the external shocks of the numerical model presented here),
then the flux is proportional to the volume of the emitting region
($\propto r^3$) and the emissivity ($\rho B \propto \rho_{\rm ext}^{3/2}$), and
will also depend on the characteristic spectral break frequencies
$\nu_m$ and $\nu_c$ \citep[see for example][]{2010ApJ...722..235V}:

\begin{equation} F \propto r^3 \rho_{ext}^{3/2} \nu_m^{(p-1)/2}
\nu_c^{1/2}, \end{equation}

where

\begin{equation} \nu_m \propto \rho_{ext}^{1/2},~~ \nu_c \propto
\rho_{ext}^{-3/2} t^{-2},~~ r \propto t. \end{equation}

The flux as a function of time is therefore:

\begin{equation} F \propto t^2 \rho_{ext}^{(p+2)/4} = t^{2-k(p+2)/4}.
\end{equation}

The shape of the plateau in this model is then $t^{-\alpha_X}$, where
$\alpha_X = -2 + k(p+2)/4$.  For $k=2$, $\alpha_X = (p-2)/2$, which is
always shallow and decaying for reasonable values of $p$.  For $k = 2$
and $p = 2.5$, this gives $\alpha_X = 1/4$ (This slope is shown in
Figure \ref{fig:afterglow}).  It is also straightforward to see from
this that a negative slope favors $k = 2$ (For example, a uniform
external density profile with $k = 0$ would always produce a rising
afterglow slope, rather than a plateau).

Thus, the environment within $\sim 10^{16}$ cm of the progenitor must be
a wind ($k \approx 2$) in order for this model to make sense \hilight{(though as mentioned in Section 2.1, the wind density in this study corresponds to $A = A_*$ ($5 \times 10^{11}$ g/cm$^2$), where $A_* = 24$, larger than typical inferred values).}  Additionally, the model predicts a particular slope to the spectrum, $\nu^{-p/2}$.  Analysis of these slopes in afterglow data has been carried out by \cite{2012ApJ...744...36S}, who claim that a majority (55\%) of bursts are consistent with this model.

\subsection{Prediction of Steeper Slopes at Lower Frequencies}

Optical observations of afterglow plateaus have found steeper slopes,
with  $\alpha_{\rm op}$ closer to unity.  This is consistent with the
coasting model; if the frequency is below the cooling break, $\nu_m <
\nu < \nu_c$, then it is straightforward to show that $\alpha_{\rm op} = -3
+ k(p+5)/4$.  For $k=2$, this gives $\alpha_{\rm op} = (p-1)/2 = 3/4$ for $p
= 2.5$.  Independent of p, one finds the relationship

\begin{equation} \Delta \alpha \equiv \alpha_{op} - \alpha_X = 3k/4-1
\approx 1/2. \end{equation}

More precisely, this model predicts that $\Delta \alpha \approx 1/2$ or
$0$, depending on whether the optical band is below the cooling break. 
This prediction is in contrast with the $\Delta \alpha = 1/4$, $0$
prediction for a decelerating blastwave
\citep[e.g.][]{1998ApJ...497L..17S}.  Many afterglow light curves appear
to be consistent with $\Delta \alpha = 1/2$ and incompatible with
$\Delta \alpha = 1/4$ \citep{2013AnA...557A..12Z}, which may prove to be
an important confirmation of this model.

This exercise should be confirmed in detail with an accurate multi-band synchrotron model.  In particular, it will be important to determine whether this model can account for a missing break in the optical band at the end of the plateau, which appears to be the case in many observations.

\hilight{An additional calculation of an optical afterglow is shown in Figure \ref{fig:xrayopt}, but for this set of model parameters, the optical frequency is also above the cooling break and therefore has the same plateau slope as in the X-rays.  At early times, however, the steep decay has a shallower slope in the optical than in the X-rays.}

\subsection{Prediction of a Microwave Flash}

The characteristic synchrotron frequency of the reverse shock scales as $\nu_m \propto \Gamma^4$ \citep{1999ApJ...520..641S}.  If the Lorentz factor of the afterglow jet is $\Gamma \sim 30$ rather
than $\Gamma \sim 300$, this means that reverse shock emission is
prominent at much lower frequencies than previously thought, in the
microwave rather than optical band.  Optical reverse shock emission
might then be negligible when compared to the forward shock emission, in
contrast with modeling of GRB 990123, for example
\citep{1999ApJ...517L.109S}.

\subsection{Very Steep Decay of Internal Shock Emission}

The extremely steep decay at early times ($\sim t^{-8}$, seen in Figure
\ref{fig:afterglow}) is difficult to explain for a relativistic flow,
because any steep decay in emission will be smoothed by the different
arrival times of photons coming from different angles.  However, if the jet's properties vary significantly with opening angle, it is possible that this effect is compensated for.  \hilight{The large right-most panel} of Figure \ref{fig:pretty} shows the jet core as it is about to collide with the outer shell.  The shape of the jet core suggests that the vary last part of the core to collide will be a narrow tail at low latitude $\theta \ll \theta_j$.  The very last part of the internal shock emission is then dominated by \hilight{emission from $\theta \ll \theta_j$}, so that photons from larger opening angles might not smooth the light curve significantly.

%One possible
%reason that this steep decay is possible for the current model is that
%the internal shocks may be planar while the flow at late times is
%radially directed.  This is difficult to be sure of, as the internal
%shocks are not resolved well enough to clearly distinguish between
%planar and radial flow, but this is be a possible explanation. 
%High-resolution three-dimensional studies focusing specifically on this
%internal shock production could help to shed light on this.}

\subsection{Measurement of the Electron Distribution}

If one assumes the density profile is in fact a wind ($k = 2$), this
provides a means to measure $p$, the slope of the electron power
spectrum in the shock.  Choosing $k = 2$ gives $p = 2 + 2 \alpha_X$. 
Many X-ray plateau slopes cluster around $\alpha_X \sim 0.2$, favoring
values of $p \sim 2.4$.  However, this is a highly simplified reading of the data.  For example, it is assumed here that there is no time-dependence to the shock microphysical parameters.  The slopes are also time-dependent, as the light curve transitions from one regime to another, and it is not clear where these slopes should be measured from in order to achieve a clear interpretation.

\subsection{Measurement of the Lorentz Factor}

The deceleration time in a wind is given by $t_{\rm decel} \sim t_{\rm Sedov} /
\Gamma^2$.  In observer time this is $t^{\rm obs}_{\rm decel} \sim
t_{\rm Sedov}/\Gamma^4$ (The Sedov time $t_{\rm Sedov} \equiv E_{\text{\rm iso}} /
(\rho_{\rm wind} R_3^2 c^3)$ is defined here to be the time when the Jet has
swept up a rest mass comparable to its isotropic equivalent energy). 
Typical GRB parameters ($E_{\rm iso} \sim 10^{54}$ erg, $\rho_{\rm wind} R_3^2 \sim 10^{13}$ g/cm) can give $t_{\rm Sedov} \sim 10^{10}$ seconds, and
the break occurs at $t^{\rm obs}_{decel} \sim 10^4$ seconds, which strongly
constrains typical Lorentz factors during the coasting phase to be
$\Gamma \sim 30$.  Such strong constraints may be very powerful for fitting afterglow data with initial jet models.

The plateau duration can be estimated as

\begin{equation}
t_{\rm decel} / t_{\rm shocks} \sim ( t_{\rm Sedov} / \tau_0 ) \Gamma^{-4} 
\end{equation}

\begin{equation}
\sim { t_{\rm Sedov} / 10^{10} ~ \text{sec} \over \tau_0 / 10 ~ \text{sec} } \left( 100 \over \Gamma \right)^4.
\end{equation}

This is consistent with the current results, where the plateau persists over a few orders of magnitude, so that $t_{\rm decel}/t_{\rm shocks} \sim 100$.

This motivates numerical studies exploring the parameter space of
coasting jets with modest Lorentz factors, which should enable good fits
of these model parameters to afterglow data, analogous to what has
already been carried out for decelerating initial jet models
\citep{2014arXiv1405.5516R, 2014arXiv1405.4867Z}.

%\subsection{Measurement of the Engine Duration}

%During the early post-breakout phase, the jet consists of a
%baryon-loaded component with thickness $\Delta r_{shell}$ and a core
%with thickness $\sim \tau_0$.  The collision between the two components
%lasts until the entire core is completely shocked, at $t \sim \tau_0
%\Gamma^2$.  In observer time, this is $t_{\rm obs} \sim \tau_0 (1+z)$ (the
%factors of $\Gamma$ cancel, and $z$ is the redshift of the burst).  This
%provides a means of measuring the engine duration, $\tau_0 =
%t_{decay}/(1+z)$, where $t_{decay}$ is the final peak in the X-rays
%before the steep decay.  The light curve in the current model peaks at
%$t_{decay} \sim 40$ seconds, meaning if the redshift were known, one
%would infer an engine duration of $\tau_0 \sim 12$ seconds, which is
%fairly accurate for this example.  Improvements can be made on this by
%directly modeling engines of various durations and measuring $t_{decay}$
%from the numerical output.

\subsection{Measurement of the Progenitor Mass}

If the Lorentz factor $\Gamma$ can be constrained by the deceleration
break time, this has important implications for understanding
progenitors; if the energy $E$ and Lorentz factor $\Gamma$ are known,
then $M = E/\Gamma$ is the mass loaded into the jet from the cocoon.  It
is possible that this mass could help constrain the mass of the
progenitor, possibly using analytical models for the jet and cocoon
\citep[e.g.][]{2011ApJ...740..100B, 2014arXiv1402.4142B} to connect the
stellar mass to the ejecta mass.  Such models would need to be revised to include the emergence of this massive shell in front of the jet core.

Alternatively, this question could be explored numerically, by
performing high-resolution three-dimensional calculations of jets
breaking out of various stellar models like those of
\cite{2004ApJ...608..365Z}, to determine if there is any reliable
relationship between ejecta mass and progenitor mass.

\subsection{Implications for Prompt Emission}

Internal shocks have long been a leading model for prompt emission,
though they usually are thought to result from the collision of multiple
highly relativistic shells.  If the jet is top-heavy such that the shell
in front is baryon-loaded (as proposed here), this causes the collision
to be much more violent, and should increase the efficiency of internal
shocks substantially.  However, this picture does not preclude other models for prompt emission.  It is possible that the steep decay is produced by internal shocks, but that some other mechanism is responsible for the gamma rays themselves.  Internal shock models may have difficulty explaining the details of the variability in the prompt emission.

\subsection{Implications for Short Bursts}

Everything invoked here applies to the collapsar scenario
\citep{1993ApJ...405..273W, 1999ApJ...524..262M, 2001ApJ...550..410M}, which is thought to
describe long GRBs.  The afterglows of short bursts may therefore be
quite different.  If there is no cocoon then jets from short bursts may
have very little baryon loading.  The prompt emission might also be
produced by some other mechanism.  Additionally, the environment
surrounding short bursts may be very different from a $k=2$ wind, so
even if a coasting phase is detected, this phase might be characterized
by a rise instead of a plateau.

Some short bursts are followed by a long tail of extended emission
\citep{2006ApJ...643..266N, 2009ApJ...696.1871P} which might be
interpreted as an afterglow \hilight{(However, extended emission can exhibit rapid variability, which is not accounted for in this model)}.  If the environment of the burst is uniform ($k=0$), then instead of a plateau one would expect a rise $\propto
t^2$, followed by a decay consistent with Blandford-McKee.  The peak of
the light curve would occur at the deceleration time,

\begin{equation} t_{\text{peak}} \sim ( E_{\text{iso}} /
\rho_{\text{ISM}} ~ c^5 ~ \Gamma^8 )^{1/3}. \end{equation}

The examples of GRB 050724 and others \citep{2014MNRAS.438..240G} have
extended emission with peak times of $\sim 100$ seconds.  Under this
interpretation, assuming $E_{\text{iso}} = 10^{51}$ erg and
$\rho_{\text{ISM}} = 10^{-24}$ g/$\text{cm}^3$, one would infer a jet
Lorentz factor \\

\begin{equation} \Gamma \sim \left( {E_{\text{iso}} \over
\rho_{\text{ISM}} ~ c^5 ~ t_{\text{peak}}^3 }\right)^{1/8} \approx 100.
\end{equation}

This large inferred Lorentz factor would suggest that either the core initially has a much larger Lorentz factor, $\gamma_{core} \sim 10^3$, before colliding with an outer shell and decelerating to $\Gamma \sim 100$, or that short bursts simply produce jets which are not top-heavy.

\section{Summary}

Collapsars naturally produce top-heavy outflows; a highly relativistic
jet core is associated with the clean outflow escaping via the
passageway tunneled out by the engine.  A less relativistic component is
associated with the baryon-loaded outer shell of the jet which was
originally in front of the jet head at breakout.

A collision between these two flows can produce a flare (possibly prompt
emission) at early observer times, followed by a steep decay in the
afterglow until $t_{\rm obs} \sim 100$ s, as seen in \textit{Swift} light
curves.  The jet core is relativistic enough to produce the prompt
emission with $\gamma_{\rm core} \sim 100$, but the late afterglow is
produced by a much less relativistic jet, formed from the amalgamation
of the shell and core.  Because of the lower Lorentz factor ($\Gamma
\sim 30$) of the less relativistic afterglow, the deceleration time of
the jet occurs at the relatively late observer time $t_{\rm obs} \sim
10^{4-5}$ seconds, and during the time in-between internal shocks and
deceleration, the afterglow can exhibit a long plateau, as seen in
Figure \ref{fig:afterglow}.

This model represents both a natural mechanism for the steep decay seen at early times, and an alternative to the late-time energy injection
model to explain plateaus in GRB afterglow light curves.  This model is purely hydrodynamic (magnetic fields are neglected), suggesting that hydro models may be accurate for most of the jet's propagation through the star, consistent with recent analysis \citep{2014arXiv1402.4142B}.  Beyond the usual collapsar scenario, no new ingredients are imposed in this model; top-heavy jets naturally arise when the jet collimation is facilitated by a stellar interior.

\acknowledgments This research was supported in part by NASA through
Chandra grant TM3-14005X and Fermi grant NNX13AO93G.

Resources supporting this work were provided by the NASA High-End
Computing (HEC) Program through the NASA Advanced Supercomputing (NAS)
Division at Ames Research Center.  We are grateful to Hendrik van
Eerten, Andrei Gruzinov and Eliot Quataert for helpful comments and
discussions.  We would also like to thank the anonymous referee for his or her complete and thorough review.

\bibliographystyle{apj} 
%\bibliography{jetbib}

\end{document}